\documentclass{statsoc} 
\usepackage{graphicx} 
\usepackage{amsmath,amssymb,verbatim} 
\usepackage{wrapfig} 
\usepackage[displaymath,pagewise]{lineno}

\newcommand{\R} {\mathbb{R}}

\newcounter{theorem}
\newtheorem{algorithm}[theorem]{Algorithm} 
 
\usepackage{graphics} 
\usepackage{natbib}

\title{Parameter Estimation for Partially Observed Hypoelliptic Diffusions} 

\author{Yvo Pokern}
  \address{ 
  Department of Statistics, University of Warwick, Coventry CV4 7AL, England. }
  \email{pokern@maths.warwick.ac.uk.} 
\author{Andrew M. Stuart}
  \address{ 
  Mathematics Institute, University of Warwick, Coventry CV4 7AL, England. }
  \email{stuart@maths.warwick.ac.uk.} 
\author[Pokern, Stuart, Wiberg]{Petter Wiberg}
  \address{Goldman-Sachs, London.}

\begin{document} 
\nolinenumbers
\maketitle 
 
\begin{abstract} 
Hypoelliptic diffusion processes can be used to model a variety of phenomena  
in applications ranging 
from molecular dynamics to audio signal analysis. We study  
parameter estimation for such processes in situations where  
we observe some components of the solution at discrete times.  
Since exact likelihoods 
for the transition densities are typically not known, approximations are used that 
are expected to work well in the limit of small inter-sample times $\Delta t$ and 
large total observation times $N\Delta t$. Hypoellipticity together with partial 
observation leads to ill-conditioning requiring a judicious combination of 
approximate likelihoods for the various parameters to be estimated. We combine these 
in a deterministic scan Gibbs sampler alternating  
between missing data in the unobserved solution components, and parameters. 
Numerical experiments illustrate asymptotic consistency of the method 
when applied to simulated data. The paper concludes with application 
of the Gibbs sampler to molecular dynamics data. 
\end{abstract} 
 
\section{Introduction}

In many application areas it is of interest to model some components 
of a large deterministic system by a low dimensional stochastic 
model. In some of these applications, insight from the deterministic 
problem itself forces structure on the form of the stochastic model, 
and this structure must be reflected in parameter estimation. In this 
paper, we study the fitting of stochastic differential equations 
(SDEs) to discrete time series data in situations where the model is a 
hypoelliptic diffusion process, meaning that the covariance 
matrix of the noise is degenerate, but the probability densities are smooth, 
and also where observations are only 
made of variables that are not directly forced by white noise. Such a 
structure arises naturally in a number of applications. 
 
One application is the modelling of macro-molecular systems, see 
\cite{GruT94} and \cite{Hum05}.  In its basic form, molecular dynamics describes
the molecule by a large Hamiltonian system of ordinary differential equations (ODEs).
\linelabel{1_7}As is commonplace in chemistry and physics, we will
refer to data obtained from numerical simulation of such models
as molecular dynamics data. If the 
molecule spends most of its time in a small number of macroscopic 
configurations then it may be appropriate to model the dynamics within, 
and in some cases between, these 
states by a hypoelliptic diffusion. While this phrasing of the question is  
relatively recent, under the name of the "Kramers problem" it dates back to  
\cite{Kra40} with a brief summary 
in section 5.3.6a of \cite{Gar85}. Another application, 
audio signal analysis, is referred to 
in \cite{God99} where a continuous time ARMA model is used, see also
\linelabel{citeGod06}\cite{God06} for more on the type of methodology used.

We consider SDE models of the form 
\begin{linenomath}
\begin{eqnarray} \label{eq:Langevin} 
 \left\{\begin{array}{rcl} \mathrm{d}x &=& \Theta A(x) \mathrm{d}t + C \mathrm{d}B\\ 
  x(0) &=& x_0 \end{array}\right. 
\end{eqnarray} 
\end{linenomath}
where $B$ is an $m$-dimensional Wiener process and 
$x$ a $k$-dimensional continuous process with $k>m$.  
$A:\mathbb{R}^k \longrightarrow \mathbb{R}^l$ is a set of 
(possibly non-linear) globally Lipschitz force functions. 
The parameters which we estimate are the last $m$ rows of the drift  matrix
(the first $k-m$ rows of which are assumed to be known),  
$\Theta \in \R^{k \times l}$, and 
the diffusivity matrix $C$ which we assume to be of the form 
\begin{linenomath}\begin{eqnarray*} 
C &=& \left[ \begin{array}{c} 0 \\ \Gamma  \end{array} \right] \in 
\R^{k\times m} 
\end{eqnarray*}
\end{linenomath}
where $\Gamma \in \R^{m \times m}$ is a constant nonsingular matrix.  
Thus, we are estimating drift and diffusion parameters only in the coordinates
which are directly driven by white noise.

It is known that under suitable hypotheses on $A$ and $C$, a unique  
$L^2$-integrable solution $x(\cdot)$ exists almost-surely 
for all times $t\in\R^+$, see e.g. Theorem 5.2.1 in \cite{Oksendal00}.
We also assume that the process defined by (\ref{eq:Langevin}) is hypoelliptic
as defined in \cite{Nualart91}.
Intuitively, this corresponds to the noise being spread 
into all components of the system \eqref{eq:Langevin} via the drift.  

The structure of $C$ implies that the noise acts directly only on a subset of 
the variables which we refer to as {\em rough}.  
It may then be transmitted, through the coupling in the drift, 
to the remaining parts of the system which we refer to as  
{\em smooth} (we do not mean $C^\infty$ here, but they are at least $C^1$). 
To distinguish between rough and smooth variables, we introduce 
the notation $x(t)^T =(u(t)^T,v(t)^T)$ where $u(t)\in \R^{k-m}$ is smooth and 
$v(t) \in \R^m$ is rough. 
It is helpful to define projections  
$\mathcal{P}:\R^k\rightarrow \R^{k-m}$ by $\mathcal{P}x=u$ and  
$\mathcal{Q}:\R^k\rightarrow \R^m$ by 
$\mathcal{Q}x=v$. 
 
We denote the sample path at $N+1$ equally spaced 
points in time by $\{x_n=x(n \Delta t)\}_{n=0}^N$, and we write $x_n^T 
=(u_n^T, v_n^T)$ to separate the rough and smooth components. Also, for any sequence 
$(z_1,\ldots, z_N), \, N \in \mathbb{N}$ we write 
$\Delta z_n = z_{n+1}-z_n$ to denote forward differences. 
We are mainly interested in cases where only the smooth component, $u$, is observed 
and  our 
focus is on parameter estimation for all of  
$\Gamma$ and for entries of those rows of $\Theta$ corresponding 
to the rough path, on the 
assumption that $\{u_n\}_{n=0}^N$ are samples from a true solution of  
(\ref{eq:Langevin}); such a parameter estimation problem arises naturally 
in many applications and an example is given in section \ref{sec:butane}.
We will describe a deterministic scan Gibbs sampler to approach this problem,
sampling alternatingly from the missing path $\{v_n\}_{n=0}^N$, the drift parameters
$\Theta$ and the covariance $\Gamma\Gamma^T$. 
It is natural to consider $N\Delta t=T \gg 1$ and $\Delta t \ll 1$. 

Given prior distributions for the parameters, $p_0(\Theta,\Gamma\Gamma^T)$, 
the posterior distribution can be constructed as follows: 
\begin{linenomath}
\begin{equation}\label{eq:likelihoodIntro} 
\left. \begin{array}{rcl}  
\mathbb{P}(v,\Theta,\Gamma\Gamma^T | u) &=& 
\frac{\mathbb{P}(v,\Theta,\Gamma\Gamma^T,u)}{\mathbb{P}(u)} \\ 
&\propto & \mathcal{L}(u,v|\Theta,\Gamma\Gamma^T) \frac{p_0(\Theta, \Gamma\Gamma^T)}{\mathbb{P}(u)} 
\end{array}\right.  
\end{equation} 
\end{linenomath}
Here, $\mathcal{L}(u,v|\Theta,\Gamma\Gamma^T)$ has been introduced as a measure
equal to the probability density $\mathbb{P}(u,v|\Theta,\Gamma\Gamma^T)$ up to
a constant of proportionality. \linelabel{3^14}When $u,v$ are fixed and $\mathcal{L}(u,v|\Theta,\Gamma\Gamma^T)$
is thought of as a function of $\Theta$ and $\Gamma\Gamma^T$ it is a
a likelihood.
 
Similarly, the probability densities $\mathbb{P}(v|\Theta,\Gamma\Gamma^T,u)$,
$\mathbb{P}(\Theta|v,\Gamma\Gamma^T,u)$ and $\mathbb{P}(\Gamma\Gamma^T|v,\Theta,u)$
are replaced by corresponding expressions using $\mathcal{L}$ when omitting constants
of proportionality that are irrelevant to estimation of the posterior 
probability.
The probability density $\mathbb{P}(u,v|\Theta,\Gamma\Gamma^T)$ 
gives rise to the transition density $\mathbb{P}(u_{n+1},v_{n+1}|u_n,v_n,\Theta,\Gamma\Gamma^T)$
which we will write as $\mathcal{L}(u_{n+1},v_{n+1}|u_n,v_n,\Theta,\Gamma\Gamma^T)$
when omitting constants of proportionality. 

In principle, \eqref{eq:likelihoodIntro} can be used as the basis for Bayesian sampling of 
$(\Theta, \Gamma\Gamma^T)$, viewing $v$ as missing data. 
However, the exact probability of the path, $\mathbb{P}(u,v|\Theta, \Gamma\Gamma^T)$,   
is typically unavailable. In this paper we will combine judicious 
approximations of this density to solve the sampling problem. 

The sequence $\{x_n\}_{n=0}^N$ defined above is generated by a Markov chain.  
The random map $x_n \mapsto x_{n+1}$ is determined by the integral equation
\begin{linenomath}
\begin{eqnarray*}
x_{n+1} &=& x_n + \int_{n \Delta t}^{(n+1)\Delta t} \Theta A(x(s)) ds + 
\int_{n\Delta t}^{(n+1)\Delta t}C dB(s).
\end{eqnarray*}
\end{linenomath}
The Euler-Maruyama approximation of this map gives
\begin{linenomath}
\begin{eqnarray}
\label{eq:EMapprox}
X_{n+1} &\approx& X_n + \Delta t \Theta A(X_n) + \sqrt{\Delta t} R(0,\Theta)\xi_n
\end{eqnarray}
\end{linenomath}
where $X_n,\xi_n \in \mathbb{R}^k$, $\xi_n$ is an iid sequence of 
normally distributed random variables, $\xi_n\sim \mathcal{N}(0,I)$, and
\begin{linenomath}
\begin{eqnarray*}
R(0,\Theta) = \begin{bmatrix} 0 & 0 \\ 0 & \Gamma \end{bmatrix} \in \mathbb{R}^{k\times k}
\end{eqnarray*}
\end{linenomath}
is not invertible. (\linelabel{3_1}Here, as throughout, we use uppercase letters to denote discrete-time approximations
of the continuous time process.)
This approximation corresponds to retaining the terms of order
$\mathcal{O}(\Delta t)$ in the drift and of $\mathcal{O}(\sqrt{\Delta t})$ in the
noise when performing an Ito-Taylor expansion (see chapter 5 of \cite{Kloeden92}). 
Due to the non-invertibility of
$R(0,\Theta)$, this approximation is unsuitable for many purposes and we
extend it by adding the first non-zero noise terms arising in the first 
$k-m$ rows of the It\^o-Taylor expansion for $X_{n+1}$. This results in the expression
\begin{linenomath}
\begin{equation} \label{eq:StatModel1}  
X_{n+1}  \approx  X_n + \Delta t \Theta A(X_n) + \sqrt{\Delta t} R(\Delta t; \Theta) \xi_n 
\end{equation}
\end{linenomath}
where $X_n \in \mathbb{R}^k, \, \xi_n \in \mathbb{R}^k$ is distributed as  
$\mathcal{N}(0,I)$ and $R(\Delta t; \Theta) \in \mathbb{R}^{k \times k}$. 
Because of the hypoellipticity, $R(\Delta t; \Theta)$ is now invertible, but  
the zeros in $C$ mean that it is highly 
ill-conditioned (or near-degenerate) for $0 < \Delta t \ll 1$. Specific
examples for the matrix $R$ will be given later.

\linelabel{3_12}Ideally we would like to implement the following deterministic scan Gibbs sampler:
\vspace{1ex} 
\begin{enumerate} 
\item 
Sample $\Theta$ from $\mathbb{P}(\Theta| u,v,\Gamma\Gamma^T)$. 
\item
Sample $\Gamma\Gamma^T$ from $\mathbb{P}(\Gamma\Gamma^T| u,v,\Theta)$. 
\item
Sample $v$ from $\mathbb{P}(v | u,\Theta, \Gamma\Gamma^T)$. 
\item 
Restart from step (a) unless sufficiently equilibrated. 
\end{enumerate} 
\vspace{1ex} 
In practice, however, approximations to the densities $\mathbb{P}$ will be needed.
We refer to expressions of the form \eqref{eq:StatModel1} as \linelabel{3_13} models 
and we will use them to approximate the exact density on path-space, 
$\mathbb{P}(u,v|\Theta, \Gamma\Gamma^T)$, of the path $u,v$ for 
parameter values $\Theta$ and $\Gamma\Gamma^T$.  
\linelabel{3_10}The resulting approximations,
$\mathbb{P}_E(U,V|\Theta,\Gamma\Gamma^T)$ and $\mathbb{P}_{IT}(U,V|\Theta,\Gamma\Gamma^T)$ 
of $\mathbb{P}(u,v|\Theta,\Gamma\Gamma^T)$, 
are found from \eqref{eq:EMapprox} and \eqref{eq:StatModel1} respectively.
We again use $\mathcal{L}_E$ and $\mathcal{L}_{IT}$ 
in the same way as for the exact distribution $\mathbb{P}$ above when omitting 
constants of proportionality.

The questions we address in this paper are: 
 
\linelabel{3_6}\begin{enumerate} 
\item[i] 
How does the ill-conditioning of the Markov chain $\{x_n\}_{n=0}^N$  
affect parameter estimation for $\Gamma\Gamma^T$ and for the last $m$  
rows of $\Theta$ in the regime  
$\Delta t \ll 1, \, N\Delta t = T \gg 1$ ?  
\item[ii] 
In many applications, it is natural that only the smooth data $\{u_n\}_{n=0}^N $ 
is observed, and not the rough data $\{v_n\}_{n=0}^N$. What effect does  
the absence of observations of the rough data have on the estimation for  
$\Delta t \ll 1$ and $N\Delta t=T \gg 1$? 
\item[iii]
The exact likelihood is usually not available; what approximations of the  
likelihood should be used, in view of the ill-conditioning? 
\item[iv]
How should the answers to these questions be combined to produce 
an effective Gibbs loop to sample the distribution of parameters $\Theta, \, \Gamma\Gamma^T$ 
and the missing data $\{v_n\}_{n=0}^N$?  
\end{enumerate}  
 
To tackle these issues, we use a combination of analysis and 
numerical simulation, based on three model problems which are conceived  
to highlight issues central to the questions above.  
We will use analysis to explain why 
some seemingly reasonable methods fail, and simulation will be used 
both to extend the validity of the analysis and to illustrate good 
behaviour of the new method we introduce.   
 
For the numerical simulations, we will use either exact discrete time 
samples of \eqref{eq:Langevin} in simple Gaussian cases, or  
trajectories obtained by Euler-Maruyama simulation of the SDE on a 
temporal grid with a spacing considerably finer 
than the observation time interval $\Delta t$. 

In section 2 we will introduce our three model problems and in section 3 we  
study the performance of $\mathcal{L}_E$  
to estimate the diffusion coefficient. Observing and analysing 
its failure in the case with partial observation leads to the improved statistical 
model yielding $\mathcal{L}_{IT}$ which eliminates these problems; 
we introduce this in section 4. In section 5 we show that $\mathcal{L}_{IT}$  
is inappropriate for drift estimation, but that $\mathcal{L}_E$ is effective in this 
context.  
In section 6, the individual estimators will be combined into a Gibbs sampler 
to solve the overall 
estimation problem with asymptotically consistent performance being demonstrated 
numerically. Section 7 contains an application to molecular 
dynamics and section 8 provides concluding discussion. 

We introduce one item of notation to simplify the presentation.  
Given an invertible matrix $R\in\mathbb{R}^{n\times n}$ 
we introduce a new norm using the Euclidean norm on $\mathbb{R}^n$ by setting 
$\|x\|_R=\|R^{-1}x\|_2$ for vectors $x\in \mathbb{R}^n$. 

\subsection{Two classical estimators} 
\linelabel{4_14}From previous work on hypoelliptic diffusions, we note a classical
estimator for the covariance matrix and for the drift matrix in 
the linear fully observed case
which will be useful for reference later in the paper. 

Firstly, it is straightforward to estimate the covariance matrix 
$\Gamma\Gamma^T$ from the quadratic variation:
noting that 
\begin{linenomath}
\begin{equation} \label{eq:diffusionCoefficient} 
  \frac{1}{T} \sum_{n=0}^{N-1} (v_{n+1}-v_n)(v_{n+1}-v_n)^T  
  \to \Gamma \Gamma^T \quad 
  \text{as} \quad N \to \infty,
\end{equation} 
\end{linenomath}
with $T=N \Delta t$ fixed, see \cite{Durrett96}.

The Girsanov formula gives rise to a maximum likelihood estimator for
the lower rows of $\Theta$, and 
in the linear case, where $A$ is just the identity, 
the maximum likelihood estimate for the whole of $\Theta$ is given by 
\begin{linenomath}
\begin{equation} \label{eq:LeBretonDrift} 
{\hat \Theta}=[\int_0^T dx x^T][ \int_0^T xx^T dt]^{-1}. 
\end{equation} 
\end{linenomath}
For the hypoelliptic case,
this is proved to be consistent as $T \to \infty$ in \cite{LeB85}.

\section{Model Problems} 
To study the performance of parameter estimators, we have 
selected a sequence of three Model Problems ranging from simple 
linear stochastic growth through a linear oscillator subject to noise and damping  
to a nonlinear oscillator of similar form. All these problems are second order
hypoelliptic and they have a physical
background, so we use $q$ (position) and $p$ (momentum) to denote smooth and rough components
in the Model problems instead of $u$ and $v$ which we used in the general case.  
Their general form is given as the second order Langevin equation  
\begin{linenomath}\begin{equation}\label{eq:OscFramework} 
\left\{ \begin{array}{rcl} 
dq &=& p dt, \\ 
dp &=& \left( -\gamma p + f(q;D) \right) dt + \sigma dB 
\end{array}\right. 
\end{equation}\end{linenomath} 
where $f$ is some (possibly nonlinear) force-function parametrised by $D$ and the variables 
$q$ and $p$ are scalar. The parameters $\gamma$, $D$ and $\sigma$ are to be estimated.

\subsection{Model Problem I: Stochastic Growth} 
Here, $x=(q,p)^T$ satisfies 
\begin{linenomath}\begin{equation} \label{eq:stochasticGrowth} 
  \left\{ \begin{array}{rcl} 
    dq &=& p dt \\ 
    dp &=& \sigma dB. 
  \end{array} \right. 
\end{equation}\end{linenomath} 
The process has one parameter, the diffusion parameter $\sigma$, that 
describes the size of the fluctuations.  In the setting of 
\eqref{eq:Langevin} we have 
\begin{linenomath}\[ 
  A(x) = x 
  \quad , \quad 
  \Theta = \begin{bmatrix} 0 & 1 \\ 0 & 0 \end{bmatrix}, 
  \quad  
   C = \begin{bmatrix} 0 \\ \sigma \end{bmatrix} 
\]\end{linenomath} 
and $u=q$, $v=p$.  The process is Gaussian with mean and covariance  
\begin{linenomath}\[ 
  \mu(t) 
  = \begin{bmatrix} 1 & t \\ 0 & 1 \end{bmatrix} 
  \begin{bmatrix} q_0 \\ r_0 \end{bmatrix} 
  \qquad \text{and} \qquad 
  \Sigma(t) 
  = \sigma^2  
  \begin{bmatrix} t^3/3 & t^2/2 \\ t^2/2 & t \end{bmatrix}. 
\]\end{linenomath} 
The exact discrete samples may be written as 
\begin{linenomath}\begin{equation} \label{eq:stochasticGrowthExact} 
  \left\{\begin{array}{rcl} 
     q_{n+1} &=& q_n + p_n \Delta t 
      + \sigma \frac{(\Delta t)^{3/2}}{\sqrt{12}} \zeta_n^{(1)} 
      + \sigma \frac{(\Delta t)^{3/2}}{2} \zeta_n^{(2)}, \\ 
     p_{n+1} &=& p_n + \sigma \sqrt{\Delta t} \zeta_n^{(2)}, 
  \end{array} \right. 
\end{equation}\end{linenomath} 
with  $\zeta_0 \sim \mathcal{N}(0,\begin{bmatrix} 1 & 0 \\ 0 & 1 \end{bmatrix})$ and  
$\{\zeta_n\}_{n=0}^N$ being i.i.d.; individual components of $\zeta_n$ are referred to
as $\zeta^{(1)}_n$ and $\zeta^{(2)}_n$ respectively. The matrix $R$ from 
\eqref{eq:StatModel1} is given here as
\begin{linenomath}\[
R=\sigma \begin{bmatrix}\frac{1}{\sqrt{12}} \Delta t & \frac{1}{2} \Delta t \\
0 & 1\end{bmatrix}.
\]\end{linenomath}
In the case of this model problem, the \linelabel{3_13_2}auxiliary model \eqref{eq:StatModel1} 
is actually exact.
 
\subsection{Model Problem II: Harmonic Oscillator} 
 
As our second model problem we consider a damped harmonic oscillator  
driven by a white noise forcing where $x=(q,p)^T$: 
 
\begin{linenomath}\begin{equation}\label{eq:harmOsc} 
\left\{ \begin{array}{rcl}  
dq &=& p dt\\ 
dp &=& -D q dt - \gamma p dt + \sigma dB. 
\end{array}\right. 
\end{equation}\end{linenomath} 
This model is obtained from the general SDE (\ref{eq:Langevin}) for the choice 
\begin{linenomath}\[ 
A\left(x \right) = x, \quad  
\Theta=\begin{bmatrix} 0 & 1 \\ -D & -\gamma \end{bmatrix},  \quad  
C = \begin{bmatrix} 0 \\ \sigma\end{bmatrix}   
\]\end{linenomath} 
and $u=q$, $v=p$.  
The process is Gaussian and the mean and covariance of the solution can be  
explicitly calculated. The matrix $R$ is the same as in Model Problem I. 
 
\subsection{Model Problem III: Oscillator with Trigonometric Potential} 
 
In the third model problem, $x=(q,p)^T$ describes the dynamics of a  
particle moving in a potential which is a  
superposition of trigonometric functions and in contact with a heat bath  
obeying the fluctuation-dissipation relation, see \cite{Las94}. This potential  
is sometimes used in molecular dynamics in  connection with the dynamics of  
dihedral angles -- see section \ref{sec:butane}. The model is  
\begin{linenomath}\begin{equation} \label{eq:TrigOsc} 
  \left\{\begin{array}{rcl} 
     dq &=& pdt, \\ 
     dp &=& (- \gamma p - \sum_{j=1}^c D_j \sin(q)\cos^{j-1}(q))dt +  \sigma dB. \\ 
  \end{array}\right. 
\end{equation}\end{linenomath} 
This equation has parameters $\gamma$, $D_i,\, i=1,\ldots,c$ and $\sigma$. It can 
be obtained from the general SDE \eqref{eq:Langevin} for the choice 
\begin{linenomath}\[ 
A\left(\begin{bmatrix} q \\ p \end{bmatrix} \right) =  
\begin{bmatrix}  \sin(q)\\ \sin(q)cos(q) \\ \vdots \\ \sin(q)\cos^{c-1}(q) \\ p  \end{bmatrix},  
\quad   
\Theta=\begin{bmatrix} 0 & \ldots & 0 & 1 \\ -D_1 & \ldots & - D_c  & -\gamma \end{bmatrix},  
 \quad  
C = \begin{bmatrix} 0 \\ \sigma\end{bmatrix}   
\]\end{linenomath} 
and $u=q$, $v=p$. 
No explicit closed-form expression for the solution of the SDE  
is known in this case; the process is not Gaussian. The matrix $R$ in the statistical
model \eqref{eq:StatModel1} is the same as the one obtained in Model Problem I.
 
\section{Euler Auxiliary Model}\label{s:EulerModel} 

\linelabel{7_14}As discussed in the introduction, we need to find appropriate
approximations for $\mathbb{P}$ in steps (a)--(c) of the desired
Gibbs loop. The purpose of this section is to show that use of
$\mathbb{P}_E$ in step (c), to sample the missing component of
the path, leads to incorrect estimation of the diffusion coefficient.
The root cause is the numerical differentiation for the missing path
which is implied by the Euler approximation.

\subsection{Auxiliary Model} 
If the force function $A(\cdot)$ is nonlinear, closed-form expressions  
for the transition density are in general unavailable. To overcome this obstacle, one 
can use a discrete time \linelabel{3_13_3}auxiliary model. The Euler model \eqref{eq:EMapprox}
is commonly used and we apply it to a simple linear model problem 
to highlight its deficiencies in the case of partially observed data from 
hypoelliptic diffusions.  
 
The Euler-Maruyama approximation of the SDE (\ref{eq:Langevin}) is 
\begin{eqnarray}\label{eq:eulerModel} 
X_{n+1} &=& X_n + \Delta t \Theta A(X_n) + \sqrt{\Delta t} C \xi_n 
\end{eqnarray} 
where $\xi_n \sim \mathcal{N}(0,I)$ is an i.i.d. sequence of $m$-dimensional 
vectors with standard normal distribution. This corresponds to \eqref{eq:StatModel1}  
with $R(\Delta t; \Theta)$ replaced by $R(0;\Theta)$ from \eqref{eq:EMapprox}. 
Thus we obtain 
\begin{linenomath}\begin{equation} 
\left\{ \begin{array}{rcl} 
U_{n+1} &=& U_n + \Delta t \mathcal{P} \Theta A(X_n) \\ 
V_{n+1} &=& V_n +\Delta t \mathcal{Q} \Theta A(X_n) + \sqrt{\Delta t} \Gamma \xi_n 
\end{array}\right\} 
\end{equation}\end{linenomath}  
where now each element of the i.i.d. sequence $\xi_n$ is distributed as $\mathcal{N}(0,I)$ 
in $\mathbb{R}^m$.  
This model gives rise to the following density: 
\begin{linenomath}\begin{equation}\label{eq:NDLikely} 
\begin{array}{l} 
\mathcal{L}_{ND}(U,V| \Theta, \Gamma\Gamma^T) = \\ \prod_{n=0}^{N-1}  
\frac{\exp\left( - \frac{1}{2}  
\left\|  \Delta V_n- \Delta t \mathcal{Q}  \Theta A(X_n) \right\|_\Gamma^2\right)} 
{\sqrt{2 \pi |\Gamma \Gamma^T|}} 
\delta \left( \frac{U_{n+1}-U_n}{\Delta t} -\mathcal{P} \Theta A(X_n)\right) 
\end{array}. 
\end{equation}\end{linenomath} 
The Dirac mass insists that the data is compatible with the \linelabel{3_13_4}auxiliary model
(\ref{eq:eulerModel}),
i.e. the $V$ path must be given by numerical differentiation (ND) of the 
$U$ path in the case of (\ref{eq:OscFramework}), and similar formulae in the general case.
To estimate parameters we will use the following expression:
\begin{linenomath}\begin{equation}\label{eq:eulerLikely} 
\begin{array}{l} 
\mathcal{L}_E(U,V| \Theta, \Gamma\Gamma^T) =  \prod_{n=0}^{N-1}  
\frac{\exp\left( - \frac{1}{2}  
\left\|  \Delta V_n- \Delta t \mathcal{Q}  \Theta A(X_n) \right\|_\Gamma^2\right)} 
{\sqrt{2 \pi |\Gamma \Gamma^T|}} 
\end{array},.
\end{equation}\end{linenomath} 
\linelabel{8^10}In the case when the Euler model is used to estimate missing components
we assume that $\{U_n\}, \, \{V_n\}$ are related so that the
data is compatible with the \linelabel{3_13_5}auxiliary model -- that is, numerical
differentiation is used to find $\{V_n\}$ from $\{U_n\}$.

\subsection{Model Problem I} 
 
The Euler \linelabel{3_13_6}auxiliary model for this model problem is 
\begin{linenomath}\begin{equation} \label{eq:stochasticGrowthEuler} 
  \begin{cases} 
    & Q_{n+1} =  Q_n + P_n \Delta t, \\ 
    & P_{n+1} =  P_n + \sigma \sqrt{\Delta t} \xi_n. 
  \end{cases} 
\end{equation}\end{linenomath} 
Here, $\{\xi_n\}$ is an i.i.d. ${\cal N}(0,1)$ sequence. 
The root cause of the phenomena that we discuss in this paper is 
manifest in comparing \eqref{eq:stochasticGrowthExact} and 
\eqref{eq:stochasticGrowthEuler}. The difference is that the  
$O((\Delta t)^{3/2})$ white noise contributions in the exact time 
series \eqref{eq:stochasticGrowthExact} do not appear in the equation for 
$Q_n$. We will see that this plays havoc with parameter estimation, 
even though the Euler method is path-wise convergent. 
 
We assume that observations of the smooth component only, $Q_n$,  
are available. In this case the Euler method for estimation  
(\ref{eq:stochasticGrowthEuler}) gives the formula 
\begin{linenomath}\begin{equation}\label{eq:numDiff} 
  P_n = \frac{Q_{n+1}-Q_n}{\Delta t}  
\end{equation}\end{linenomath} 
for the missing data.  
In the following numerical experiment we generate exact data from  
\eqref{eq:stochasticGrowthExact} using the parameter value $\sigma=1$.  
We substitute $P_n$ given by \eqref{eq:numDiff} into  
\eqref{eq:eulerLikely} and find the maximum likelihood estimator for $\sigma$  
in the case of partial observation. In the case of complete observation 
we use the exact value for $\{P_n\}$, from \eqref{eq:stochasticGrowthExact}, 
and again use a 
maximum likelihood estimator for $\sigma$ from \eqref{eq:eulerLikely}. 
 
\begin{figure} 
\begin{center} 
\includegraphics[scale=0.26]{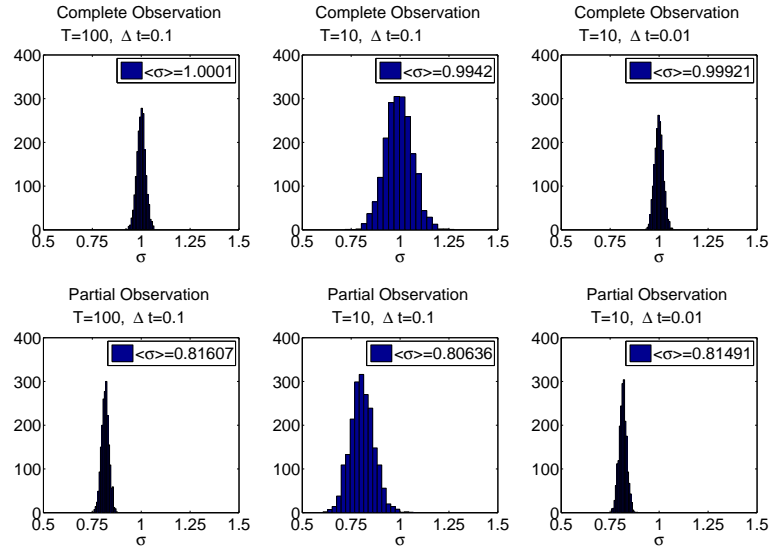}  
\end{center} 
\begin{tabular}{rcl} 
\hspace{1cm} & 
\begin{minipage}{9.5cm} 
\caption{Maximum likelihood estimates of $\sigma$ using Euler Model for Model Problem I.\newline 
Top row: fully observed process; bottom row: partially observed process.} 
\end{minipage} 
&  
\end{tabular} 
\label{fig:StochasticGrowth} 
\end{figure}

Using $N=100$ timesteps for a final time of $T=10$ with $\sigma=1$ the  
histograms for the estimated diffusion coefficient  
presented in the middle column of Figure \ref{fig:StochasticGrowth} 
are obtained. The top row contains 
histograms obtained in the case of complete observation where good agreement 
between the true $\sigma$ and the estimates is observed. The bottom row 
contains the histograms obtained for partial observation using  
\eqref{eq:numDiff}. The observed mean value of $\mathbb{E} \widehat{\sigma}=0.806$  
indicates that the method yields biased estimates.  
Increasing the final time to $T=100$ (see left column of graphs in 
Figure \ref{fig:StochasticGrowth}) or 
increasing the resolution to $\Delta t=0.01$ (see right column of graphs in Figure
\ref{fig:StochasticGrowth}) do not remove this bias. 
 
Thus we see that, in the case of partial observation, $\widehat{\sigma}$ 
contains $O(1)$ errors 
which do not diminish with decreasing $\Delta t$ and/or increasing 
$T=N\Delta t.$ 
 
\subsection{Analysis of why the missing data method fails}\label{ss:EulerMissingAnalysis} 
 
Model Problem I can be used to illustrate why this method fails. We 
first argue that the method works without hidden data. Interpreting 
\eqref{eq:eulerLikely} as a log-likelihood function wrt. $\sigma$, we obtain
following expression in the case of stochastic growth: 
\begin{linenomath}\[ 
  \log \mathcal{L}_E (\sigma|Q,P) 
  = -2 N \log \sigma - \frac{1}{\sigma^2 \Delta t}  
  \sum_{n=0}^{N-1} (\Delta P_{n})^2 
\]\end{linenomath} 
where $\Delta$ is the forward difference operator. The maximum of the 
log-likelihood function gives the maximum likelihood estimate, 
\begin{linenomath}\begin{equation} \label{E:diffusion estimate} 
  \widehat{\sigma}^2 = \frac{1}{N \Delta t} \sum_{n=0}^{N-1} (\Delta P_n)^2. 
\end{equation}\end{linenomath} 
In the case of complete data, \eqref{eq:stochasticGrowthExact} gives 
\begin{linenomath}\begin{equation} \label{E: stoch growth, diffusion MLE} 
  \widehat{\sigma}^2 = \frac{\sigma^2}{N} \sum_{n=0}^{N-1} (\zeta_n^{(2)})^2. 
\end{equation}\end{linenomath} 
By the law of large numbers, $\widehat{\sigma}^2 \to \sigma^2$ almost 
surely as $N \to \infty$.  This shows that the method works when  
the complete data is observed. 
 
Let us consider what happens when \linelabel{10^13}$P$ is hidden. In this case, $P_n$ 
is estimated by 
\begin{linenomath}\[ 
  \widehat{P}_n = \frac{Q_{n+1}-Q_n}{\Delta t}. 
\]\end{linenomath} 
But since $q_n$ is generated by \eqref{eq:stochasticGrowthExact} we find 
that 
\begin{linenomath}\[ 
  \widehat{P}_n = \frac{P_{n+1}+P_n}{2}  
  + \sigma \frac{\sqrt{\Delta t}}{\sqrt{12}} \zeta_n^{(1)} 
\]\end{linenomath} 
and \begin{linenomath}
\begin{align*} 
  \Delta \widehat{P}_n &= \frac{\Delta P_{n+1}}{2}  
  + \frac{\Delta P_{n}}{2} 
  + \sigma \frac{\sqrt{\Delta t}}{\sqrt{12}}  
    \left( \zeta_{n+1}^{(1)}-\zeta_{n}^{(1)} \right) \\ 
  &=\frac{\sigma \sqrt{\Delta t}}{2} \left( \zeta_{n+1}^{(2)}  
     + \zeta_n^{(2)}  
     + \frac{1}{\sqrt{3}} \zeta_{n+1}^{(1)}  
     - \frac{1}{\sqrt{3}} \zeta_{n}^{(1)} \right) 
\end{align*} \end{linenomath}
When $\Delta \widehat{P}_n$ is inserted in \eqref{E:diffusion 
estimate} it follows that 
\begin{linenomath}\begin{align*}
  \widehat{\sigma}^2  
  &= \frac{\sigma^2}{4 N} \sum_{n=0}^{N-1} 
  \left( \zeta_{n+1}^{(2)} + \zeta_n^{(2)}  
    + \frac{\zeta_{n+1}^{(1)} - \zeta_n^{(1)}}{\sqrt{3}} \right)^2 \\ 
  &= \frac{\sigma^2}{4 N} \bigg[  
    \sum_{n=0}^{N-1}  
    \left( \zeta_{n+1}^{(2)}+\frac{\zeta_{n+1}^{(1)}}{\sqrt{3}} \right)^2 
   +\sum_{n=0}^{N-1}  
    \left( \zeta_{n}^{(2)}-\frac{\zeta_{n}^{(1)}}{\sqrt{3}} \right)^2 \\ 
  &\qquad + 2  
    \sum_{n=0}^{N-1}  
    \left( \zeta_{n}^{(2)}-\frac{\zeta_{n}^{(1)}}{\sqrt{3}} \right) 
    \left( \zeta_{n+1}^{(2)}+\frac{\zeta_{n+1}^{(1)}}{\sqrt{3}} \right) \bigg]. 
\end{align*}\end{linenomath}
The random variables $\{\zeta_n\}_{n=0}^{N}$ are i.i.d with $\zeta_0 
\sim N(0,I)$. So, by the law of large numbers, 
 $ \widehat{\sigma}^2 \to \frac{2}{3} \sigma^2$ almost surely as  
$\quad N \to \infty$. 
Furthermore, the limits hold in either of the cases where either $N 
\Delta t=T$ or $\Delta t$ are fixed as $N \to \infty$. This means that 
independently of what limit is considered, a seemingly reasonable 
estimation scheme based on Euler approximation results in $O(1)$ 
errors in the diffusion coefficient. 
There is similarity here with work of \linelabel{11^6}\cite{GaiL97} 
showing that adaptive methods for SDEs get the 
quadratic variation wrong if the adaptive strategy is not chosen 
carefully.  
 
\section{Improved Auxiliary Model} 
The Euler \linelabel{3_13_7}auxiliary model fails to propagate noise to the smooth component of the 
solution and thus leads to estimating missing paths $v$ with incorrect quadratic variation. 
A new \linelabel{3_13_8}auxiliary model is thus proposed which propagates the noise 
using what amounts to an  It\^o-Taylor expansion, 
retaining the leading order component of the noise 
in each row of the equation. The model is used to set up 
an estimator for the missing path using a Langevin sampler from path-space which 
is then simplified to a direct sampler in the Gaussian case. 
Numerical experiments indicate that the method 
yields the correct quadratic variation for the simulated missing 
path. 

The model is motivated using our common framework the Model Problems I, II  
and III, namely \eqref{eq:OscFramework}. 
The improved \linelabel{3_13_9}auxiliary model is based on the observation that in the second 
row of an It{\^o}-Taylor expansion of  
\eqref{eq:OscFramework} the drift terms are of size 
$\mathcal{O}(\Delta t)$ whereas the random forcing term is  
"typically" (in root mean square) of size $\mathcal{O}(\sqrt{\Delta t})$.  
Thus, neglecting the contribution of the drift term in the second row on the 
first row leads to the  
following approximation of \linelabel{11_16}(\ref{eq:OscFramework}): 
\begin{linenomath}\[ 
\begin{bmatrix} 
Q_{n+1} \\ P_{n+1}  
\end{bmatrix} 
= \begin{bmatrix} 
Q_n \\ P_n \end{bmatrix} + 
\Delta t 
\begin{bmatrix}  
P_n \\ f(Q_n) -\gamma P_n  
\end{bmatrix} 
+\sigma \begin{bmatrix} 
\int_{n\Delta t}^{(n+1)\Delta t} \left(B(s) -B(n\Delta t)\right) ds \\ B((n+1)\Delta t)-B(n\Delta t) 
\end{bmatrix} 
\]\end{linenomath} 
The random vector on the right hand side is Gaussian, and can 
be expressed as a linear combination of two independent normally 
distributed Gaussian random variables. Computation of the variances and  
the correlation is straightforward leading to the following statistical 
model: 
\begin{linenomath}\begin{equation} 
\begin{bmatrix} 
Q_{n+1} \\ P_{n+1}  
\end{bmatrix} 
= \begin{bmatrix} 
Q_n \\ P_n \end{bmatrix} + 
\Delta t 
\begin{bmatrix} 
P_n \\ f(Q_n) -\gamma P_n  
\end{bmatrix} 
+\sigma \sqrt{\Delta t} R \begin{bmatrix} 
\xi_1 \\ \xi_2 \end{bmatrix} 
\label{eq:fudgeModel} 
\end{equation}\end{linenomath} 
Here, $\xi_1$ and $\xi_2$ are independent normally distributed Gaussian random 
variables and $R$ is given as 
\begin{linenomath}\begin{eqnarray*} 
R&=& \begin{bmatrix} 
\frac{\Delta t}{\sqrt{12}} & \frac{\Delta t}{2} \\ 
0 & 1 \end{bmatrix} 
\end{eqnarray*}\end{linenomath} 
 
This is a specific instance of \eqref{eq:StatModel1}. 
It should be noted that this model is in agreement with  
the Ito-Taylor approximation up to error terms of order  
$\mathcal{O}(\Delta t^2)$ in the first row and  
$\mathcal{O}(\Delta t^{\frac{3}{2}})$ in the second row
and that higher order hypoelliptic processes can be approximated
using a similarly truncated Ito-Taylor expansion. The key
important idea is to propagate noise into all components
of the system, to leading order.
 
If complete observations are available, this model performs  
satisfactorily for the estimation of $\sigma$. 
This can be verified analytically for 
Model Problem I in the same fashion as in section  
\ref{ss:EulerMissingAnalysis}.  
Numerically, this can be seen from  
the first row (referring to complete observation) 
of Figure \ref{fig:StochasticGrowthFF} for  
Model Problem I and from the first row of  
Figure \ref{fig:harmOscFF} for Model Problem II. In both cases 
the true value is given by $\sigma=1$. See subsection 
4.2 for a full discussion of these numerical experiments. 
 
If only partial observations are available, however, a means of  
reconstructing the hidden component of the path must be  
procured. A standard procedure would be the use of the K\'alm\'an 
filter/smoother (\cite{Kalman60}, \cite{Catlin89}) which could then 
be combined with the expectation-maximisation (EM) algorithm  
(\cite{DemLR77},\cite{MenD97}) to estimate parameters.  
In this paper, however, we employ a Bayesian approach sampling directly  
from the posterior 
distribution for the rough component, $p$, without factorising the sampling into 
forward and backward sweeps. 
 
\subsection{Path Sampling}\label{ss:PathSampling} 
 
The logarithm of the density on path space   
for the missing data induced by the \linelabel{3_13_10}auxiliary model \eqref{eq:StatModel1} 
can be written as follows: 
\begin{linenomath}\begin{equation}\label{eq:litDef} 
\log \mathcal{L}_{IT}(p|q,\Theta,\Gamma\Gamma^T) 
= -\frac{1}{2} \sum_{l=0}^N \left\| \Delta X_l -  
\Theta A(X_l)\Delta t\right\|_R^2 + \mathrm{const}. 
\end{equation}\end{linenomath} 
We will apply this in the case \eqref{eq:fudgeModel} which is a specific 
instance of \eqref{eq:StatModel1}. 
 
One way to sample from the density on path space, $\mathcal{L}_{IT}(P)$, 
for rough paths $\{P_i\}_{i=0}^N$ is via the Langevin equation  
(see section 6.5.2 in \cite{RobCas99}) and, in general, we expect this to be effective 
in view of the high dimensionality of $P$. Other MCMC approaches may also be used.

However, when the joint distribution of \linelabel{12_17}
$\{P_i\}_{i=1}^N$ is Gaussian it is possible to generate independent samples as follows:  
note first, that in the Gaussian case, when $\mathcal{L}_{IT}$ in (\ref{eq:litDef})
is quadratic in $P$, the derivative of $\log \mathcal{L}_{IT}$ with respect to the rough path
$P$ can be computed explicitly, which is carried out in \linelabel{12_14}\cite{PokernPhD}.
For our oscillator framework, the derivative can be expressed using  
a tridiagonal, negative definite matrix $P_\mathrm{mat}$ 
with highest order stencil $-1 \; -4 \; -1$ acting on the $P$-vector plus a  
possibly nonlinear contribution $\mathcal{Q}(Q)$ acting on the $Q$-vector only:  
\begin{linenomath}\[ 
\nabla_p \log \mathcal{L}_{IT} (Q,P)= P_\mathrm{mat} P + \mathcal{Q}(Q). 
\]\end{linenomath} 
Then, the suggested direct sampler for $P$-paths is simply: 
\begin{linenomath}\begin{equation}
P = -P_\mathrm{mat}^{-1} \mathcal{Q}(Q) + U^{-1} \xi 
\label{eq:directSampler} 
\end{equation}\end{linenomath} 
Here $U^TU=-P_\mathrm{mat}$ is a Cholesky factorisation and $\xi$ is
a dimension $N$ vector of iid normally distributed random numbers. 
 
\subsection{Estimating Diffusion Coefficient and Missing Path} 
 
The approximation $\mathcal{L}_{IT}(P,Q|\sigma, \Theta)$  
can be used to estimate both 
the missing path $p$ and the diffusion coefficient $\sigma$ for our Model Problems I, II and III. 
 
In order to estimate $\sigma$, the derivative of the logarithm of $\mathcal{L}_{IT}$
\begin{linenomath}\[\log \mathcal{L}_{IT}(\sigma|P,Q,\Theta)=\log \mathcal{L}_{IT}(P,Q|\sigma,\Theta)  
+\log {p_0(\Theta,\sigma)} +\mathrm{const}
\]\end{linenomath}  
(where priors $p_0(\Theta,\sigma)$ are assumed to be given and constants in $\sigma$
have been omitted) 
with respect to $\sigma$ is computed: 
\begin{linenomath}\begin{eqnarray*} 
\frac{\partial }{\partial \sigma} \log \mathcal{L}_{IT}&=& 
-\frac{2N}{\sigma} + \frac{1}{\sigma^3} Z + \frac{\partial}{\partial \sigma}  
\log\left( p_0(\Theta,\sigma) \right). 
\end{eqnarray*}\end{linenomath} 
Here, we have used the abbreviation 
\begin{linenomath}\begin{eqnarray*} 
Z &:=& \sum_{p=0}^{N-1} 
\left\|   
\left( 
\left(\begin{array}{c} Q_{p+1} \\ P_{p+1} \end{array} \right) 
- \left(\begin{array}{c} Q_{p} \\ P_{p} \end{array} \right) 
-\Delta t \left( \begin{array}{c}  P_p \\ 
- f(Q_n)  -\gamma P_p\end{array}\right)  
\right)\right\|^2_R. 
\end{eqnarray*}\end{linenomath} 
 
In this case no prior distribution was felt necessary 
as,  when $N\to \infty$, its importance would diminish rapidly. 
Thus we set $p_0\equiv 1$. 

We use a Langevin type sampler for this distribution. In order to avoid
the singularity at $\sigma=0$ we use the transformation $\zeta(\sigma)=\sigma^4$\linelabel{13^8}.
Using the It{\^o} formula, this yields the following Langevin 
equation which we use to sample $\zeta$ and hence $\sigma$: 
\begin{linenomath}\begin{eqnarray}\label{eq:sigmaLangevin} 
d\zeta &=& \left( (12-8N) \sqrt{\zeta} +4 Z \right) ds + 4 \sqrt{2} \zeta^\frac{3}{4} dW.
\end{eqnarray}\end{linenomath}
A simple explicit Euler-Maruyama discretisation in $s$ is used to simulate paths for 
this SDE. The timestep $\Delta s$ needs to be tuned with $N$ to ensure 
convergence of the explicit integrator. Since this is a one-dimensional problem,
conservatively small timesteps and long integration times can be afforded. With this
choice of timestep $\Delta s$ the theoretically possible transient behaviour 
(see \cite{RobTwee}) was not observed and we expect accurate samples from the posterior
in $\sigma$.
 
\begin{figure} 
\begin{center} 
\includegraphics[scale=0.37]{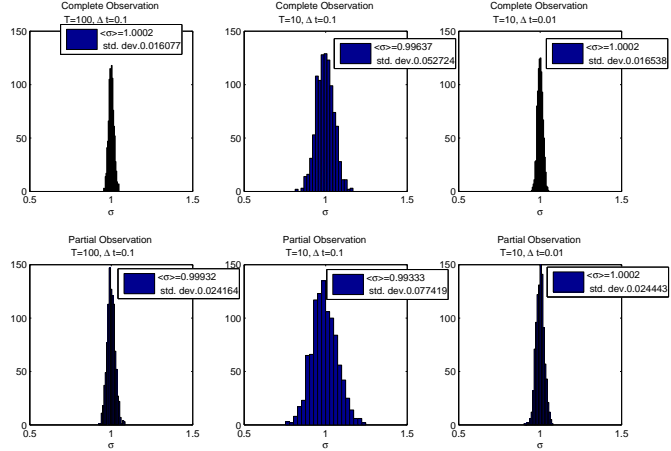}  
\end{center} 
\begin{tabular}{rcl} 
\hspace{1cm} &  
\begin{minipage}{9.5cm} 
\caption{Estimates of $\sigma$ using the $\mathcal{L}_{IT}$ Model for Model Problem I.\newline 
Top row: MLEs for fully observed process; \newline
bottom row: Mean Gibbs estimates for partially observed process.} 
\label{fig:StochasticGrowthFF} 
\end{minipage}  
& 
\end{tabular} 
\end{figure}

\begin{figure} 
\begin{center} 
\includegraphics[scale=0.37]{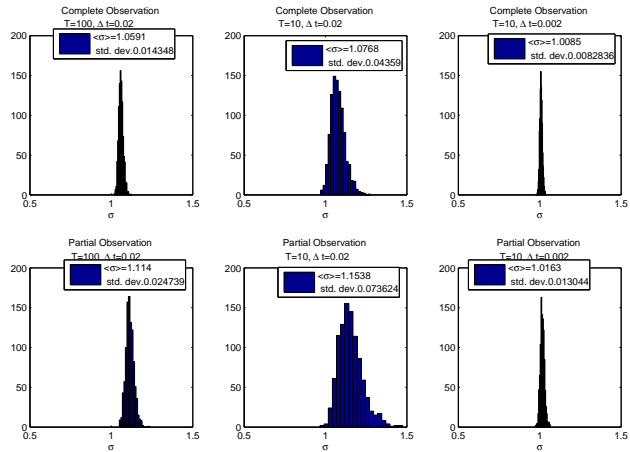}  
\end{center} 
\begin{tabular}{rcl} 
\hspace{1cm} &  
\begin{minipage}{9.5cm} 
\caption{Estimates of $\sigma$ using the $\mathcal{L}_{IT}$ Model for Model Problem II.\newline 
Top row: MLEs for fully observed process;\newline
bottom row: Mean Gibbs estimates for partially observed process.} 
\label{fig:harmOscFF} 
\end{minipage} 
& 
\end{tabular} 
\end{figure}

This Langevin-type sampler \eqref{eq:sigmaLangevin} can then be alternated in a  
Systematic Scan Gibbs Sampler  
(as described on p.130 of \cite{Liu01}) using $N_\mathrm{Gibbs}$ iterations 
with the direct sampler for the paths, 
\eqref{eq:directSampler}. This yields estimates of the missing path and the  
diffusion coefficient which is estimated by averaging over the latter half of the
$N_\mathrm{Gibbs}$ samples. We illustrate this with an example using  
Model Problem I with the following parameters: 
\begin{center} 
\vspace{1ex} 
\begin{tabular}{llll} 
$\sigma =1$, & $T\in\{10,100\}$, & $\Delta t\in\{0.1,0.01\}$, & $N_\mathrm{Gibbs}=50$. 
\end{tabular} 
\vspace{1ex} 
\end{center} 
The sample paths used for the fitting are generated using  a sub-sampled Euler-Maruyama method with temporal grid $\frac{\Delta t}{k}$ where $k=30$.
\linelabel{13_1}The resulting histogram of mean posterior estimators is given in Figure 
\ref{fig:StochasticGrowthFF} where the first row corresponds to the behaviour 
when complete observations are available and the second row corresponds to only 
the smooth component being observed and missing data being sampled according to
\eqref{eq:directSampler}. 
For Model Problem II we use the following parameters: 
\begin{center} 
\vspace{1ex} 
\begin{tabular}{lll} 
$\sigma =1$, & $D=4$, & $\gamma=0.5$, \\ 
$T\in\{10,100\}$, & $\Delta t\in\{0.02,0.002\}$, & $N_\mathrm{Gibbs}=50$. 
\end{tabular} 
\vspace{1ex} 
\end{center} 
The sample paths used for the fitting are generated as for Model Problem I and
the experimental results are given in Figure \ref{fig:harmOscFF}.

It appears from these figures that the estimator for this joint problem 
performs well for Model Problems I and II for $\Delta t$ sufficiently small  
and $T$ sufficiently large.   
A more careful investigation of the convergence properties 
is postponed to section 6 when drift estimation will be incorporated in the 
procedure.  
 
\section{Drift Estimation} 
\subsection{Overview} 
\linelabel{15_10}With the approximations $\mathcal{L}_E$ and $\mathcal{L}_{IT}$ in place, 
the question arises which of these should be used to estimate the drift  
parameters. Using Model Problem II  
we numerically observe that an 
$\mathcal{L}_E$ based maximum likelihood  
estimator performs well.  In contrast, ill-conditioning due to hypoellipticity 
leads to error amplification and 
affects the performance of the $\mathcal{L}_{IT}$ based maximum likelihood estimator.  
 
\subsection{Drift Parameters from $\mathcal{L}_E$} 
\linelabel{15_6}In order to simplify analysis, we illustrate the estimator using 
Model Problems  II, \eqref{eq:harmOsc} and III, \eqref{eq:TrigOsc}. For the latter,
the Euler \linelabel{3_13_11}auxiliary model is given as follows: 
\begin{linenomath}\begin{eqnarray} \label{eq:EulerMaru}
\left\{ \begin{array}{rcl} Q_{n+1} &=& Q_n + \Delta t P_n\\ 
P_{n+1} &=& P_n - \Delta t \sum_{i=1}^cD_if_i(Q_n) - \Delta t \gamma P_n + \sqrt{\Delta t} \sigma \xi_n  
\end{array}\right. ,
\end{eqnarray}\end{linenomath}
where we abbreviated the trigonometric expressions using $f_j(q)=\sin(q)\cos^{j-1}(q)$.
The functional $\mathcal{L}_E$ in this case is given by: 
\begin{linenomath}\begin{equation} 
\label{eq:EulerLikelihood} 
\begin{array}{rcl} 
\mathcal{L}_E (\gamma,D|Q,P,\sigma) &\propto &
 \exp \left( 
- \sum_{n=0}^{N-1}  
\frac{\left(\Delta P_{n}+\Delta t\sum_{i=1}^cD_if_i(Q_n) 
+ \Delta t \gamma P_n \right)^2}{2\Delta t \sigma^2 } \right)  
\end{array}.
\end{equation}\end{linenomath} 
\linelabel{16^1}Clearly, this posterior is Gaussian with distribution
\begin{linenomath}\begin{equation} \label{eq:ThetaLangevin} 
\widehat{\Theta} \sim \mathcal{N}\left(M_E^{-1} b_E ,M_E^{-1}\right),
\end{equation}\end{linenomath} 
where the matrix $M_E$ and the vector $b_E$ can be read off from (\ref{eq:EulerLikelihood}).

\subsection{Drift Parameters from $\mathcal{L}_{IT}$} 
 
As the approximate model based on $\mathcal{L}_{IT}$ is observed to resolve the difficulty 
with estimating $\sigma$ for hidden 
$p$-paths, it is interesting to see whether it can also be used  
to estimate the drift parameters.   
 
The logarithm of the density on path space up to an additive constant
is given by \eqref{eq:litDef}. To illustrate 
the problems arising from the use of $\mathcal{L}_{IT}$ we use Model Problem II, so  
that \eqref{eq:litDef} becomes 
\begin{linenomath}\begin{eqnarray} 
\log \mathcal{L}_{IT} (\Theta|Q,P,\sigma) &= & \frac{1}{2  \Delta t} \sum_{n=0}^{N-1} \left\| 
(\Delta X_{n}-\Delta t  \Theta A(X_n)) \right\| ^2_R + \mathrm{const} 
\label{eq:logLikely} 
\end{eqnarray} \end{linenomath}
where $R=\sigma \begin{bmatrix}\frac{\Delta t}{\sqrt{12}} & \frac{\Delta t}{2} \\ 
0 & 1 \end{bmatrix}$, irrelevant constants have been omitted and we have\linelabel{16_7} 
\begin{linenomath}\[ 
A\left(\left[ \begin{array}{c} Q_n \\ P_n \end{array}\right]\right) = 
\left[ \begin{array}{c} 
Q_n \\ P_n \end{array}\right], \quad  
\Theta  = \left[ \begin{array}{cc} 
0 &  1 \\ 
-D & -\gamma \end{array}\right]. 
\]\end{linenomath} 
In order to obtain a maximum likelihood estimator from this, we  
take the derivative  
with respect to the parameters $D $ and $\gamma$  and equate to zero.  
This yields the following linear system: 
\begin{linenomath}\begin{eqnarray} 
\begin{bmatrix} 
\sum_n Q_n^2 \Delta t & \sum_n P_n Q_n \Delta t \\ 
\sum_n P_n Q_n  \Delta t& \sum_n P_n^2\Delta t\\ 
\end{bmatrix}  
\begin{bmatrix} 
\hat{D} \\ \hat{\gamma} \end{bmatrix}  
&=& 
\begin{bmatrix} 
- \sum_n Q_n \Delta P_n\\ 
- \sum_n P_n \Delta P_n\\ 
\end{bmatrix} 
+ 
\begin{bmatrix} 
\sum_n \frac{3}{2} Q_n\left(\frac{\Delta Q_n}{\Delta t} -P_n\right) \\  
\sum_n \frac{3}{2} P_n\left( \frac{\Delta Q_n}{\Delta t} - P_n \right) 
\end{bmatrix} 
\label{eq:1stFudgeDriftEst} 
\end{eqnarray} \end{linenomath}
Comparing this linear system to the mean of the successful 
estimator (\ref{eq:ThetaLangevin}) we note the presence of  
an additional term on the right hand side.
This term leads to the failure of the above estimator.  
\linelabel{16_1}Thus, $\mathcal{L}_{IT}$ is not an appropriate approximation
for use in step (a) of the Gibbs sampler. 

\subsection{Numerical Check: Drift} 
There are two factors influencing convergence: $T$ and $\Delta t$.  
To illustrate their influence, consider the following series 
of numerical tests. All of the tests share these parameters: 
\begin{center} 
\vspace{1ex} 
\begin{tabular}{llll} 
$D=4$ & $\gamma=0.5$ & $\sigma=0.5$ & $k=30$ 
\end{tabular} 
\vspace{1ex} 
\end{center} 
Data for the tests is again generated using an Euler-Maruyama method on a  
finer temporal grid with resolution $\Delta t / k $.  
\begin{figure} 
\begin{center} 
\includegraphics[scale=0.25]{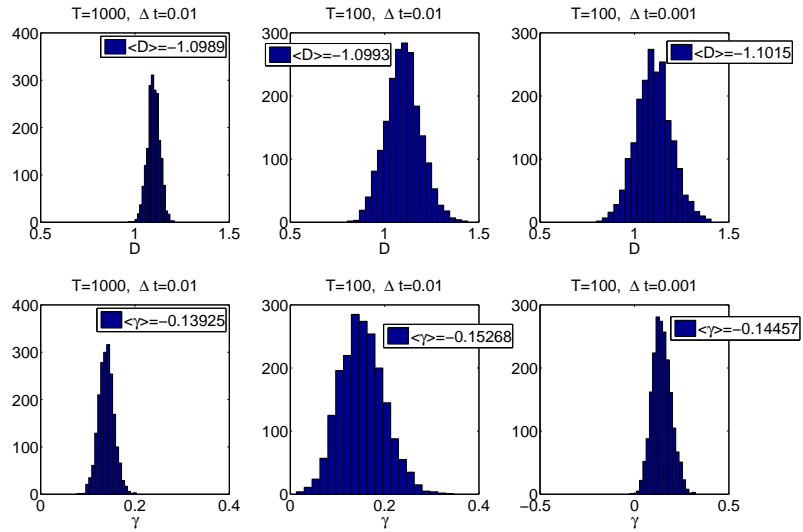} \\ 
\end{center} 
\caption{Maximum likelihood drift estimates for Model Problem II, using $\mathcal{L}_{IT}$} 
\label{fig:FirstFudgeDrift} 
\end{figure} 
In the plot given in Figure \ref{fig:FirstFudgeDrift} the top row contains 
histograms for the maximum likelihood estimate 
for the drift parameter $D$ whereas the second row contains 
histograms for the drift parameter $\gamma$ in any case using the full sample 
path for maximum likelihood inference, i.e. formula
(\ref{eq:1stFudgeDriftEst}). It is clear from these experiments summarised in Figure  
\ref{fig:FirstFudgeDrift} that both $D$ and $\gamma$ are grossly 
underestimated by $\hat{D}, \hat{\gamma}$ from (\ref{eq:1stFudgeDriftEst}).  
This problem does not resolve for smaller $\Delta t$ (see the right column of that
figure); it does not disappear for longer intervals of observation, either,
as can be inferred from the left column of Figure (\ref{fig:FirstFudgeDrift}).
 
\subsection{Why the $\mathcal{L}_{IT}$ Model Fails for the Drift Parameters} 
\linelabel{17_3}The key is to compare (\ref{eq:1stFudgeDriftEst}) with the mean 
in (\ref{eq:ThetaLangevin}). This reveals that the last term in
(\ref{eq:1stFudgeDriftEst}) is an error term which we now study.

Using the 2nd order It\^o-Taylor approximation  
\begin{linenomath}\begin{eqnarray*} 
X_{n+1} &=& X_n + \Delta t A X_n + \left[\begin{array}{ll} 
1 & 0 \\ 
-\gamma & 1  \end{array}\right] R \left[\begin{array}{l} 
\xi_1 \\ \xi_2 \end{array}\right]  
+ \frac{1}{2} \Delta t^2 A^2 X_n  
+\mathcal{O}(\Delta t^{\frac{5}{2}}) 
\end{eqnarray*}\end{linenomath} 
we can compute the second term on the right hand side of \eqref{eq:1stFudgeDriftEst}: 
\begin{linenomath}\begin{equation}\label{eq:substFFDrift} 
\begin{bmatrix} 
\sum_n \frac{3}{2} Q_n \left( \frac{\Delta Q_n}{\Delta t} - P_n \right)\\ 
\sum_n \frac{3}{2} P_n \left( \frac{\Delta Q_n}{\Delta t} - P_n \right)  
\end{bmatrix}  
= \\ 
\begin{bmatrix} 
-\frac{3}{4} \gamma \sum_n Q_n P_n \Delta t - \frac{3}{4} D \sum_n Q_n^2 \Delta t \\ 
-\frac{3}{4} D \sum_n Q_n P_n \Delta t  -  \frac{3}{4} \gamma \sum_n P_n^2 \Delta t  
\end{bmatrix} + 
I_s + \mathcal{O}(\Delta t). 
\end{equation}\end{linenomath} 
Here, $D$ and $\gamma$ refer to the exact drift parameters used to {\em generate} 
the sample path, whereas $\hat{D}$ and $\hat{\gamma}$ in \eqref{eq:1stFudgeDriftEst} 
and \eqref{eq:substFFDrift} are the drift parameters estimated using the improved  
\linelabel{3_13_12}auxiliary model. The term $I_s$ on 
the right hand side contains stochastic integrals whose expected value is zero.  
 
As the mean error terms can be written in terms of the matrix elements 
themselves, \eqref{eq:substFFDrift} can be substituted in  
\eqref{eq:1stFudgeDriftEst} to obtain: 
\begin{linenomath}\begin{eqnarray} 
\mathbb{E} \hat{D}  &=& \frac{1}{4} D + \mathcal{O}(\Delta t)\\ 
\mathbb{E} \hat{\gamma}  &=& \frac{1}{4} \gamma +\mathcal{O}(\Delta t). 
\end{eqnarray} \end{linenomath}
 
This seems to be corroborated by the numerical tests.

\subsection{Conclusion for Drift Estimation} 
 
We observed numerically but do not show here that $\mathcal{L}_E$ associated 
with an Euler model for the SDE \eqref{eq:Langevin} yields asymptotically  
consistent Langevin and maximum likelihood estimators for Model Problem II.  

While it is aesthetically desirable to base the estimation of all parameters as well as 
the missing data on the same approximation $\mathcal{L}_{IT}$  
of the true density (up to multiplicative constants) $\mathcal{L}$, and although  
this approximation was found to work well for the estimation of missing data 
and the diffusion coefficient, it does not work for the drift parameters.  
 
It is possible to trace this failure to the fact that only the second row of  
$\Theta$ is estimated where $\mathcal{O}(\Delta t)$ errors in the first row 
get amplified to $\mathcal{O}(1)$ errors in the second row. Estimating all 
entries of $\Theta$, while being outside the specification of the problem 
under consideration, also yields 
$\mathcal{O}(1)$ errors if $\mathcal{L}_{IT}$ is used and so does not remedy the problem.  
This problem is not shared by the discretised version of the  
diffusion independent estimator \eqref{eq:LeBretonDrift},  
but this is not a maximum likelihood estimator for $\mathcal{L}_{IT}$.  
 
In summary, for 
the purposes of fitting our model problems to observed data we  
employ the Euler \linelabel{3_13_13}auxiliary model \eqref{eq:EulerLikelihood} for the drift parameters. 
 
\section{The Gibbs Loop} 
\linelabel{18_1}In this section, we combine the insights obtained in previous 
sections to formulate an effective algorithm to fit hypoelliptic  
diffusions to partial observations of data at discrete times. 
We apply a deterministic scan Gibbs sampler  
alternating between missing data 
(the rough component of the path, $v$), drift parameters and diffusion parameters.  

We combine the approximations developed and motivated in previous sections
in the following Gibbs sampler:
\vspace{1ex} \linelabel{19-20}
\begin{enumerate} 
\item 
Sample $\Theta$ from  $\mathbb{P}_E(\Theta| U,V,\sigma)$. 
\item 
Sample $\sigma$ from $\mathbb{P}_{IT}(\sigma| U,V,\Theta)$. 
\item 
Sample $V$ from $\mathbb{P}_{IT}(V | U,\Theta, \sigma)$.
\item
Restart from step (a) unless sufficiently equilibrated. 
\end{enumerate} 
\vspace{1ex} 

\noindent Our numerical results will show that 
this judicious combination of approximations results in an effective
algorithm. Theoretical justification remains an interesting open problem.

When applied to Model Problem III the detailed algorithm reads as follows:
\begin{algorithm} \label{alg:Model3Alg}
Given observations $Q_i,\, i=1,\ldots,N$, the initial $P$-path is obtained 
using numerical differentiation: 
\begin{linenomath}\begin{equation}  \label{eq:NumDiffQ}
P_i^{(0)} = \frac{\Delta Q_i}{\Delta t}.
\end{equation}\end{linenomath} 
The initial drift parameter estimate is just set to zero:  
$\left\{D^{(0)}_j\right\}_{j=1}^c=0,\, \gamma^{(0)}=0$. 
Then start the Gibbs loop:\\ 

For $k=1,\ldots,N_\mathrm{Gibbs}$: 
\begin{enumerate} 
\item 
Estimate the drift parameters $\gamma^{(k)}$ and $\{D_j^{(k)}\}_{j=1}^c$  
using sampling based on $\mathcal{L}_E$ given $\left\{P_i^{(k-1)}\right\}_{i=0}^N$ via  
\eqref{eq:ThetaLangevin}. 
\item 
Estimate the diffusivity $\sigma^{(k)}$  
using the Langevin sampler \eqref{eq:sigmaLangevin}  
based on $\mathcal{L}_{IT}$ given $\left\{P^{(k-1)}_i\right\}_{i=0}^N$ and $\gamma^{(k)}$, $\left\{D^{(k)}_j\right\}_{j=1}^c$. 
\item 
Get an independent sample of the $P$-path, $\left\{P_i^{(k)}\right\}_{i=0}^N$ 
using \eqref{eq:directSampler} derived from $\mathcal{L}_{IT}$  
given parameters $\gamma^{(k)}$, $\left\{D_j^{(k)}\right\}_{j=1}^c$ and $\sigma^{(k)}$. 
\end{enumerate} 
\label{alg:Gibbs} 
\end{algorithm} 

\begin{figure} 
\begin{center} 
\includegraphics[scale=0.25]{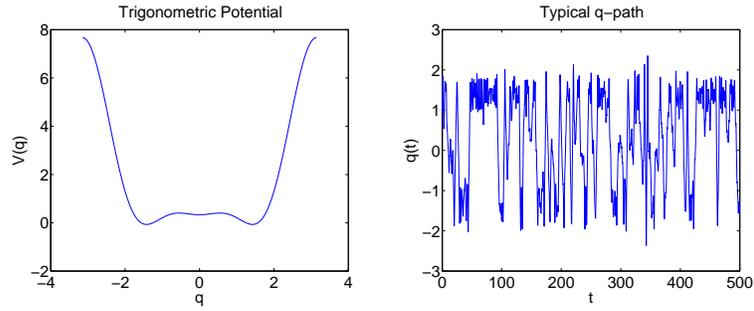} \\ 
\end{center} 
\caption{Typical sample path for Model Problem III, $T=500$} 
\label{fig:SamplePathMod3} 
\end{figure}

We test this algorithm numerically where sample paths of
\eqref{eq:TrigOsc}  are generated using 
a sub-sampled Euler-Maruyama approximation of the SDE. The data is generated  
using a timestep that is smaller than the observation time step 
by a factor of either $k=30$ or $k=60$. Comparing the results for these  
two and other non-reported cases, they are found not to depend 
on the rate of subsampling, $k$, if this is chosen large enough. The parameters used 
for these simulations are as follows: 
\begin{center} 
\vspace{1ex} 
\begin{tabular}{lllll} 
$D_0=1$, & $D_1=-8$, & $D_2=8$,  & $\gamma=0.5$, & $\sigma=0.7$,\\ 
$T=500$, & $\Delta t\in \left\{\frac{1}{2}, \ldots, \frac{1}{128}\right\}$, & 
$N_\mathrm{Gibbs}=50$. 
\end{tabular}  
\vspace{1ex} 
\end{center} 
The trigonometric potential resulting from this choice of drift parameters is 
depicted on the left of Figure \ref{fig:SamplePathMod3} and 
a typical samplepath is given on the right side of Figure \ref{fig:SamplePathMod3}. 
It should be noted that all sample paths are started at $(q,p)=(1,1)$. 
\linelabel{20_10}A typical sample path for $q$ given in Figure \ref{fig:SamplePathMod3}.

\begin{figure} 
\begin{center} 
\includegraphics[scale=0.25]{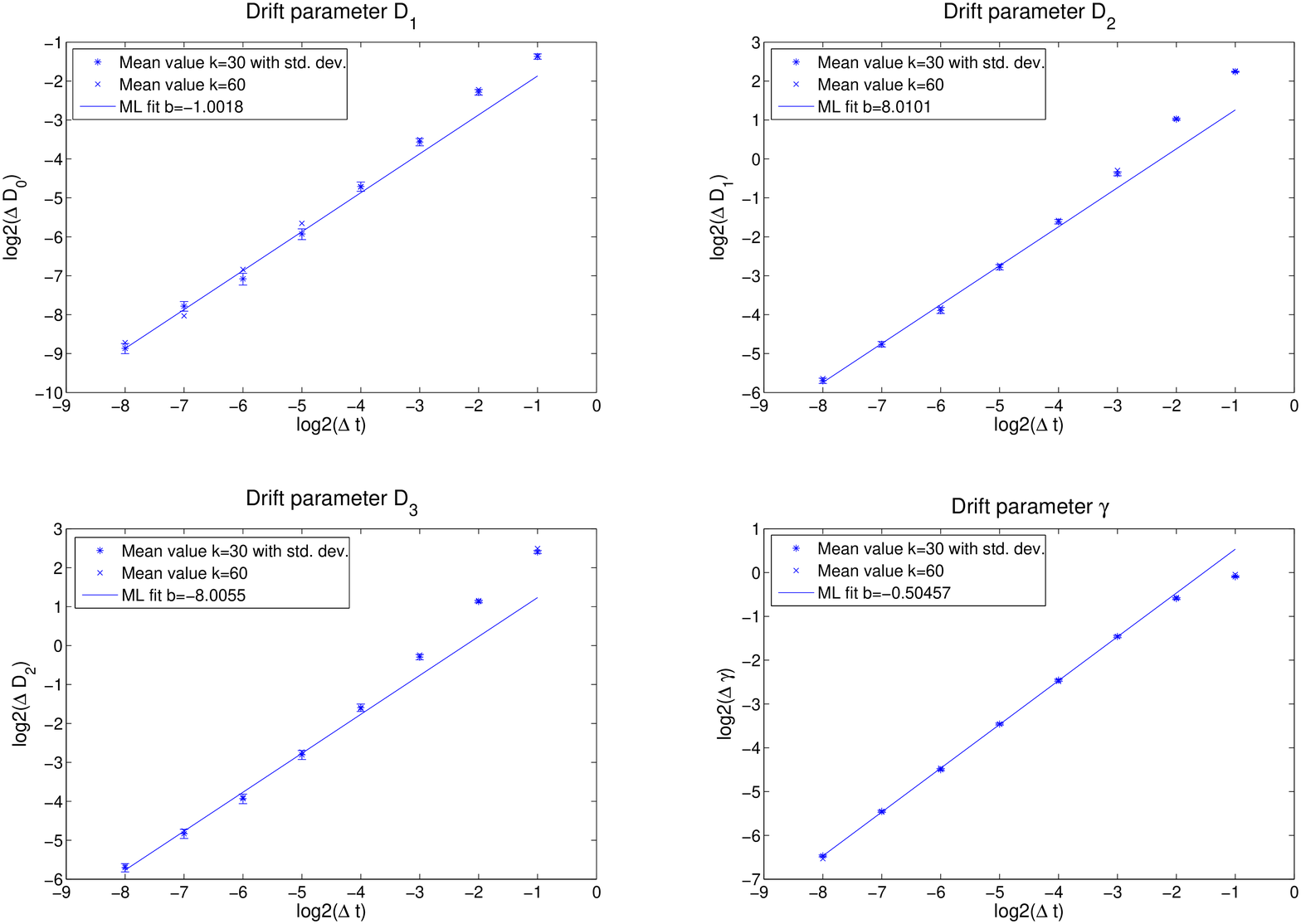} \\ 
\end{center} 
\caption{Model Problem III, $T=500$ :
Displaying Averaged Mean Posterior Deviations of the Drift Parameters} 
\label{fig:wholeLoopNonLin} 
\end{figure}

\begin{figure} 
\begin{center} 
\includegraphics[scale=0.7]{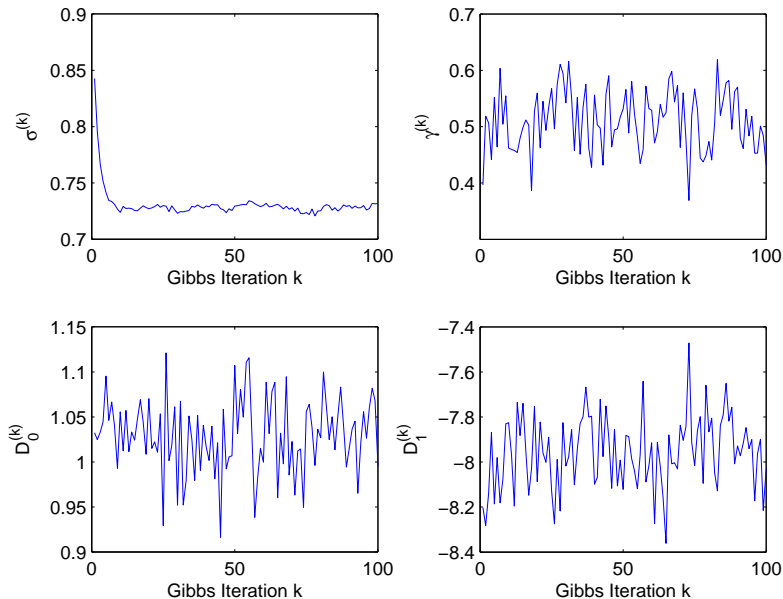} \\ 
\end{center} 
\caption{Model Problem III: Burn-in of Gibbs Sampler} 
\label{fig:GibbsBurnin} 
\end{figure}

The performance of the Gibbs sampler for the sample $q$-path given in Figure
\ref{fig:SamplePathMod3} is shown in Figure \ref{fig:GibbsBurnin}
where $100$ Gibbs steps sampling from the posterior distribution of 
drift and diffusion parameters are shown for the setup shown above except that
here $N_\mathrm{Gibbs}=100$ and $\Delta t=0.01$.
\linelabel{20_9}Mean posterior estimators are computed averaging over the latter half of $N_\mathrm{Gibbs}$
iterations as before. This sampling is repeated up to 64000 times and we label the repeated-sampling
average of these mean posterior estimators as $\langle \widehat{D_i}\rangle$ and 
$\langle \widehat{\gamma} \rangle$. We then compute their deviation from 
the true values, $\Delta D_i=\langle \widehat{D_i} \rangle -D_i $ and plot  
$\Delta D_i$ and $\Delta \gamma$ versus $\Delta t$ in a doubly logarithmic plot 
given in Figure \ref{fig:wholeLoopNonLin}.

We seek to fit a straight line to the $\Delta D_i$ in a doubly logarithmic plot
to ascertain the order of convergence. \linelabel{22^1}Since a standard least squares fit 
proves inadequate, we employ the following procedure:

Given averaged numerically observed parameter estimates 
$y_i$  and their numerically observed Monte Carlo standard deviations $\alpha_i$  
obtained at timesteps $\Delta t_i$ we fit $b$ and $c$ in the following linear regression:
\begin{linenomath}\begin{equation} 
\alpha_i \xi_i = y_i - b - c \Delta t_i.   
\label{eq:LSQfit} 
\end{equation}\end{linenomath} 
Assuming that the errors $\xi_i$ are normally distributed (which is empirically 
found to be the case) a maximum likelihood fit for the parameters $b$ and $c$ can 
be performed and yields the asymptotic (for $\Delta t \to 0$) drift parameter 
values reported in Figure \ref{fig:wholeLoopNonLin}.
Note that this fit constrains the slope of the fitted line in the doubly 
logarithmic plot to one. This is to minimise the number of parameters fitted 
and to improve the accuracy of the extrapolated value $b$ which is the 
predicted value for $y$ at $\Delta t=0$.
It can be observed in Figure \ref{fig:wholeLoopNonLin} that this leads to good 
agreement with the observed average parameter values $y_i$, and this corroborates  
the estimator's bias being of order $\mathcal{O}(\Delta t)$.
 
Comparing the results for the two final times tested, $T=50$ and $T=500$,  
we find that the deviation of the asymptotic drift parameter ($b$ in  
\eqref{eq:LSQfit}) from the true parameter value is consistent with it being 
$\mathcal{O}\left(\frac{1}{T} \right)$. This error is attributed to all sample paths  
having been started at $(q,p)=(1,1)$ rather than from a point sampled from 
the equilibrium measure.
 
\begin{figure} 
\begin{center} 
\includegraphics[scale=0.35]{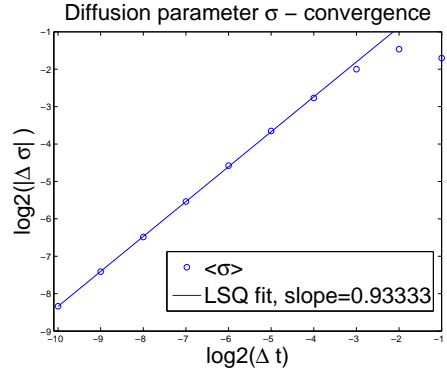} \\ 
\end{center} 
\caption{Model Problem III, $T=500$ :
Displaying Averaged Mean Posterior Deviations for $\sigma$} 
\label{fig:wholeLoopNonLinSigma} 
\end{figure}

For the diffusion parameter $\sigma$, results analogous to those in Figure 
\ref{fig:wholeLoopNonLin}, using the same parameter values, 
are shown in Figure \ref{fig:wholeLoopNonLinSigma} (although 
that figure displays results for $k=30$ only). Asymptotic consistency  
can be observed from this figure with a naive least squares fit yielding  
a slope of $\mathcal{O}(\Delta t ^{0.93}).$ This is consistent with an  
$\mathcal{O}(\Delta t)$ error in the estimated diffusion parameter. 
 
From these considerations it is apparent that the numerical experiments' outcome is 
consistent with  an  
$\mathcal{O}(\Delta t) + \mathcal{O}\left(\frac{1}{T} \right)$  
bias, so the Algorithm 
\ref{alg:Gibbs} is numerically observed to be
an asymptotically unbiased estimator of the drift and diffusion 
parameters in the cases studied. 
 
\section{Application to Molecular Conformational Dynamics} 
\label{sec:butane} 
As an application of fitting hypoelliptic diffusions using partial 
observations we consider data arising from molecular dynamics simulations
of a Butane molecule using a simple heat bath approximation.\linelabel{23^4}
 
By considering the origin of the data we demonstrate that 
it is natural to fit a hypoelliptic diffusion 
process which yields convergent results for diminishing inter-sample intervals 
$\Delta t$. Also, stabilisation of the fitted force function  
$f(q)=\sum_{j=1}^c D_j f_j(q)$ as the number of terms to be included, $c$, increases, 
is observed. Thus the Algorithm \ref{alg:Model3Alg} is shown to 
be effective on \linelabel{23^12}molecular dynamics data. It is also clear, though, that the  
resulting fit has only limited predictive capabilities as it fails 
to fit the invariant measure of the data at all well. However, this is  
a {\em modelling} issue  which is not central to this paper.

\subsection{Molecular Dynamics} 
\begin{figure}
\begin{center} 
\includegraphics[scale=0.6]{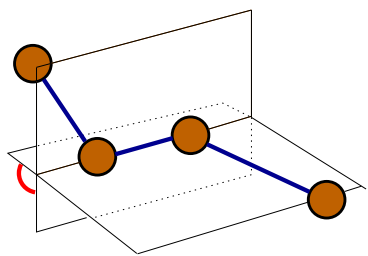}\\ 
\end{center}
\caption{Sketch of Dihedral Angle} 
\label{fig:DihedralSketch} 
\end{figure}

The data used for this fitting example are generated using a molecular dynamics (MD) 
simulation for a single molecule of Butane. In order to avoid explicit 
computations for solvent molecules, several {\em ad hoc} approximate algorithms 
have been developed in molecular dynamics. One of the more sweeping approximations 
that is nonetheless fairly popular, at least as long as electrostatic effects 
of the solvent can be neglected or treated otherwise, is Langevin dynamics.
\linelabel{23}Here, the time evolution of the Cartesian coordinates of the 
four extended atoms of Butane (see Figure \ref{fig:DihedralSketch}) is
simulated using a damped-driven Hamiltonian system; details of the force field used 
can be found in \cite{Brooks83}.

From a chemical point of view interest is focused on the dihedral angle $omega$,
which is the angle between the two planes in $\mathbb{R}^3$ formed by
atoms $1,2,3$ and atoms $2,3,4$ respectively; see the sketch
in Figure \ref{fig:DihedralSketch}. Conformational change is manifest
in this angle, and the Cartesian coordinates themselves are of little
direct chemical interest. Hence it is natural to try and describe the
stochastic dynamics of the dihedral angle in a self-contained fashion.
 
\begin{figure} 
\begin{center} 
\includegraphics[scale=0.25]{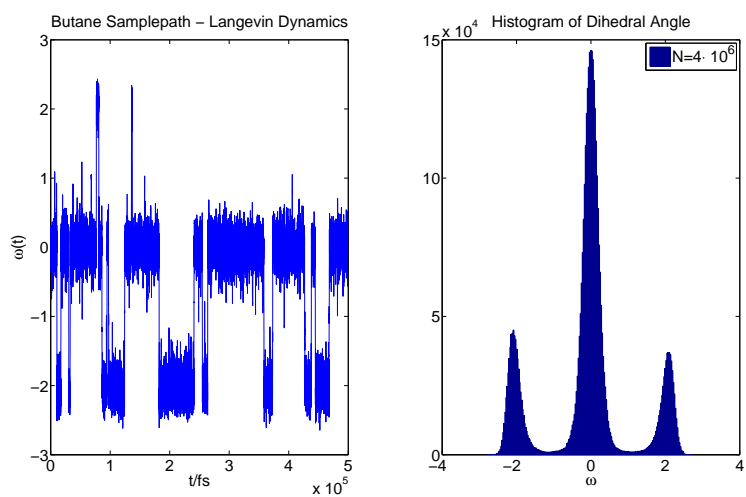} \\ 
\end{center} 
\begin{tabular}{rcl} 
\hspace{1cm} &  
\begin{minipage}{9.5cm} 
\caption{MD Samplepath Butane. \newline Left: First 500ps of sample path, 
Right: Histogram of whole sample path}
\label{fig:ButPath} 
\end{minipage}  
& 
\end{tabular} 
\end{figure} 
 
One MD run is produced using a timestep of $\Delta t=0.1$fs \linelabel{24_1}(Throughout this section,
we use the time unit femtosecond abbreviated to fs. Note that $1fs = 10^{-15}s$.)
and a Verlet variant (see p.435 in \cite{SchlickBook}) covering a total time of  
$T=4\cdot 10^{-9}$s ($4$ nanoseconds).
A section of the path of the dihedral angle as a function of time can be seen on the left of 
Figure \ref{fig:ButPath}; the corresponding histogram for the whole of the path 
is depicted to the right of that figure. 

It should be stressed that the It\^o process 
governing the behaviour of the dihedral angle $\omega$ is {\em not} of the form 
\eqref{eq:TrigOsc}, in particular, it will have a non-constant diffusivity
$\sigma$. So, fitting to this data tests the robustness of the fitting
algorithm in a way that the experiments in previous sections did not.

\subsection{Fitting} 
We aim to fit the process from Model Problem III, equation (\ref{eq:TrigOsc}),
to a subsampled trajectory of $\omega(t_i)$ \linelabel{24_11}(viewed as the smooth component $q$)
obtained from the molecular dynamics
simulation described previously. \linelabel{24_6}Subsampling is performed because we have a profusion
of data and because the hypoelliptic diffusion is expected to be a good fit only at
some timescales.

The simulation used to obtain the dihedral angle data is such that
$\omega(t)$ will be a $C^1$ function of time assuming a
suitable interpretation of the periodicity in $\omega$, so it is natural to
fit a hypoelliptic process of damped-driven Hamiltonian form.

\begin{figure} 
\begin{center} 
\includegraphics[scale=0.25]{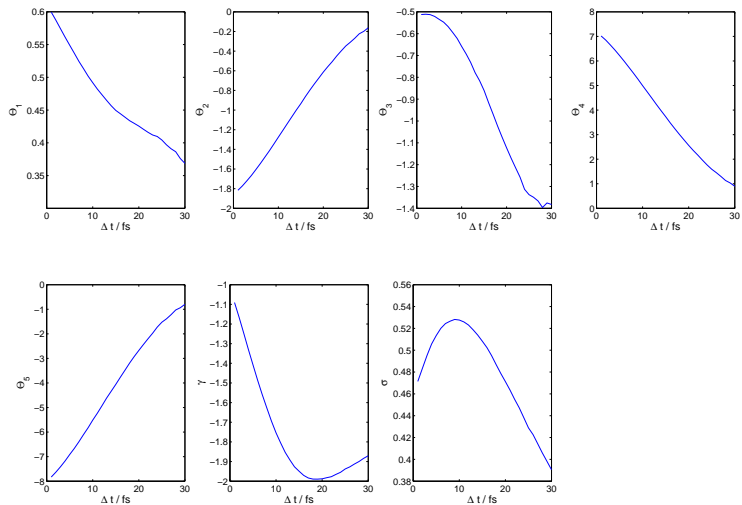} \\ 
\end{center} 
\caption{Convergence for fitted MD path with subsampling\newline
Displays mean Gibbs estimates of drift and diffusion parameters as a function
of subsampling interval $\Delta t$} 
\label{fig:ButaneDriftConv} 
\end{figure}

The physical time-units in seconds are minuscule and do not lead to  
estimated SDE parameters of order one.
It transpires that, in order to obtain parameter values of order one,
re-scaling time so that the final time  becomes $T=80000$ is a good choice.
This rescaling is useful in comparing convergence properties with what was 
observed in Section 6. 
In order to 
assess consistency, the MD data is subsampled, at timesteps  
$\Delta t \in \{1 \mathrm{fs}, \, 2  \mathrm{fs},\, 
3\mathrm{fs}\,\ldots\}$ in physical 
time units, corresponding to $\{k 0.02\}_{k\in\mathbb{N}}$ in the 
rescaled time units. Algorithm \ref{alg:Model3Alg}
is then run for $N_\mathrm{Gibbs}=40$ outer iterations on each path using a
potential ansatz 
\begin{linenomath}\begin{eqnarray*} 
V(\omega; \Theta) &=& \sum_{k=1}^c \Theta_k cos^k(\omega)
\end{eqnarray*}\end{linenomath} 
which corresponds to the force functions in (\ref{eq:TrigOsc})
setting  $D_k=k \Theta_k$ and $f=V'$; the values .
$c \in\{3,5,7\}$ are used in the sequel.
These periodic
ansatz functions are a natural choice for dihedral angle potentials, in fact,
the dihedral angle potential given in \cite{Brooks83} is of this form. 
The obtained \linelabel{25_7}drift parameter estimates under
subsampling at timestep $\Delta t$ can be seen from Figure \ref{fig:ButaneDriftConv} 
in the case $c=5$. In this figure, the sampling timestep $\Delta t$ is 
the abscissa and the \linelabel{25_5}drift and diffusion parameter estimates 
($\Theta_1,\ldots, \Theta_5$, $\gamma$ and $\sigma$)
obtained from fitting to the samplepath subsampled
at timestep $\Delta t$ is shown as the ordinate.
This plot shows the behaviour of the drift and diffusion parameter estimates averaged over
$N_\mathrm{Gibbs}=100$ Monte-Carlo samples
$\theta_1,\ldots,\theta_5,\gamma$ for different values of the subsampling rate.
\linelabel{25_1}The behaviour as $k \rightarrow 0$ indicates that the fitted parameter values converge
to a well-defined limit; $\sigma$ in particular varies relatively little over a large
range of subsampling rates. This suggests that the proposed algorithm is able to fit 
Model Problem III to molecular dynamics data. The fact that different (especially drift) 
parameter values are obtained at different subsampling rates indicates
limitations in the fit to Model Problem III and this
will be addressed in the next subsection.
 
\newpage 
\subsection{Limitations} 
The desirable convergence properties of the algorithm  in $\Delta t$ 
and $T$ should not 
be confused with inference about whether fitting this kind of model to this 
kind of MD data gives a good or a bad fit, it merely indicates that, using 
the algorithm suggested in this paper, it is possible to perform such fitting.

\begin{figure} 
\begin{center}
\includegraphics[scale=0.33]{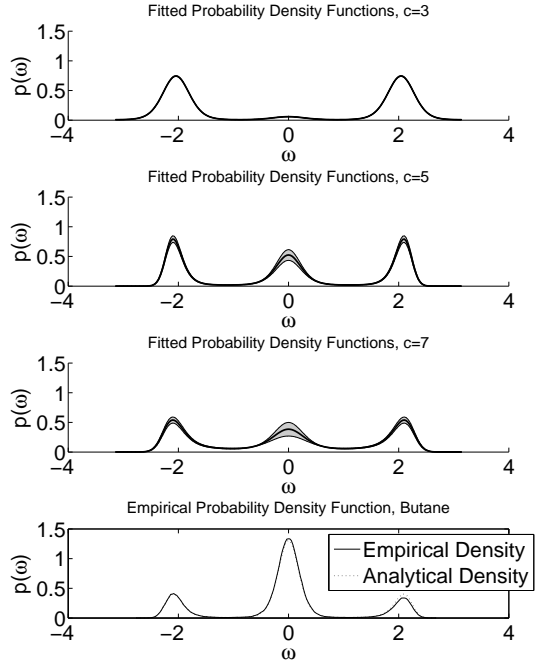} \\ 
\end{center}
\caption{PDFs resulting from fitted potentials for different  
orders of trigonometric potential - Shaded regions display posterior variance} 
\label{fig:ButaneFitHist} 
\end{figure}

To show limitations of the model in this particular application and 
see how the performance can be assessed using the fitting algorithm 
\ref{alg:Model3Alg}, we show posterior invariant  
probability densities resulting from the fitted trigonometric potentials.
In order to do this, we convert the posterior drift parameter samples 
$\{D_j^{(m)}\}_{j=1}^c$ obtained at step $m$ using input data subsampled at rate 
$k=1$ to an invariant density,
$\varrho^{(m)}$ specified by its values on an equidistant grid on the 
interval $[-\pi,\pi]$. These densities for $m\in\{1,\ldots,1000\}$ are 
then averaged and their standard deviation is computed point-wise on the grid.  
This results in the plot given in Figure \ref{fig:ButaneFitHist}.
There, we display results for three orders of trigonometric potential $c$ to 
be fitted. These are contrasted with the empirically observed invariant density and  
the density arising from the classic canonical thermodynamic ensemble which 
is proportional to $\exp\left(-\frac{V(\omega)}{kT}\right)$ \linelabel{26_14}which are given in
the plot at the bottom of Figure \ref{fig:ButaneFitHist}. For the  
\linelabel{26^16}force field used in the molecular dynamics simulation, 
it is known that the latter two agree in the limit 
$T\rightarrow \infty$, see \cite{AlexDiplom}.

\linelabel{26^11}It should be stressed that in each of these experiments, convergence diagnostics
indicate convergence of the Gibbs sampler and the posterior distributions for the
drift and diffusion parameters are very concentrated and hence posterior
variances both for the drift and diffusion parameters as well as the induced
invariant densities are low.

With increasing polynomial order $c$ we find some qualitative change in the 
resulting invariant density and also (in particular moving from $c=5$ to $c=7$) 
a marked increase in posterior variance. This goes hand-in-hand with  
a marked increase in the condition number of the drift parameter matrix  
$M_E$ in \eqref{eq:ThetaLangevin}. It is simply an ill-conditioned problem to derive 
higher and higher order polynomial coefficients from a fixed length of 
observed path. 

It is observed that even though the empirically observed invariant density is  
smooth and close to the thermodynamical expectation, the fitted potentials  
induce an SDE whose invariant measure is not a good approximation of the 
empirical density. This may simply be attributed to the fact that the SDE 
that is being fitted does not represent a good model of the {\em dynamics}  
of the dihedral angle in the Butane molecule with
second order Langevin heat bath model.
 
\section{Conclusions} 
A hybrid algorithm for fitting drift and diffusion parameters of a hypoelliptic 
diffusion process, with constant diffusivity, from observation of smooth
data at discrete times has been described. The method
combines a Gibbs sampler together with differing approximate likelihoods
employed in different steps of the Gibbs loop. Its performance has been validated 
numerically for a number of test cases and an application to molecular dynamics 
data has been given. While parameter fitting can be viewed as an inverse problem 
for SDE solvers -- and thus ill-conditioning of some kind is always to be expected -- 
a detailed understanding of the particular ill-conditioning induced by hypoellipticity
and partial observation has been attained.

While only second order hypoelliptic problems have been treated in this article,
the algorithm's applicability is expected to encompass
order $k$ hypoelliptic problems and it has been tested successfully on a third order 
example. Furthermore, non-linear $p$-dependence in the example 
\eqref{eq:OscFramework} can be dealt with using a Langevin sampler for the missing 
path and this has also been tested.
\linelabel{27_3}Additionally, observations that are not exactly equispaced can also be processed provided the
maximal inter-sample time is sufficiently small.

Further avenues of investigation include the use of imputed data-points between samples 
to diminish $\mathcal{O}(\Delta t)$ errors; however there is a risk of
bad mixing as $\sigma$ is determined by the small scale behaviour of the
process which would then be dominated by the imputed data points. This has
been analysed in the case of elliptic diffusion processes in \cite{RobS01} \linelabel{25^16}and
an application of standard estimators to this problem in the hypoelliptic case
is given in \cite{God06}.

Also, an extension to position dependent diffusion coefficients may prove
useful, in particular, in may render the algorithm more useful in
molecular dynamics contexts such as those in \cite{Hum05}.

\section*{Acknowledgements}
\linelabel{ack}The authors wish to express their gratitude towards the referees and the editor 
for their careful reading of the paper and for their constructive suggestions.
\bibliographystyle{Chicago} 
\bibliography{EstimationHypoelliptic} 

\end{document}